# Mantle Dynamics in Super-Earths: Post-Perovskite Rheology and Self-Regulation of Viscosity

[1]P. J. Tackley, [2]M. Ammann, [2]J. P. Brodholt, [2]D. P. Dobson, [3]D. Valencia

[1]Institute of Geophysics, ETH Zurich, Sonneggstrasse 5, Zurich, 8092 Switzerland (ptackley@ethz.ch; Tel: +41 446332758, Fax: +41 446331065)
[2]Department of Earth Sciences, University College London, Glower Street, London WC1E 6BT, UK
[3]Department of Earth, Atmospheric and Planetary Sciences, Massachusetts Institute of Technology, Cambridge MA 02139, USA

**Abstract**

The discovery of extra-solar "super-Earth" planets with sizes up to twice that of Earth has prompted interest in their possible lithosphere and mantle dynamics and evolution. Simple scalings suggest that super-Earths are more likely than an equivalent Earth-sized planet to be undergoing plate tectonics. Generally, viscosity and thermal conductivity increase with pressure while thermal expansivity decreases, resulting in lower convective vigor in the deep mantle, which, if extralopated to the largest super-Earths might, according to conventional thinking, result in no convection in their deep mantles due to the very low effective Rayleigh number. Here we evaluate this. First, as the mantle of a super-Earth is made mostly of post-perovskite we here extend the density functional theory (DFT) calculations of post-perovskite activation enthalpy of to a pressure of 1 TPa, for both slowest diffusion (upper-bound rheology) and fastest diffusion (lower-bound rheology) directions. Along a 1600 K adiabat the upper-bound rheology would lead to a post-perovskite layer of a very high (~$10^{30}$ Pa s) but relatively uniform viscosity, whereas the lower-bound rheology leads to a viscosity increase of ~7 orders of magnitude with depth; in both cases the deep mantle viscosity would be too high for convection. Second, we use these DFT-calculated values in numerical simulations of mantle convection and lithosphere dynamics of planets with up to ten Earth masses. The models assume a compressible mantle including depth-dependence of material properties and plastic yielding induced plate-like lithospheric behavior. Results confirm the likelihood of plate tectonics for planets with Earth-like surface conditions (temperature and water) and show a novel self-regulation of deep mantle temperature. The deep mantle is not adiabatic; instead feedback between internal heating, temperature and viscosity regulates the temperature such that the viscosity has the value needed to facilitate convective loss of the radiogenic heat, which results in a very hot perovskite layer for the upper-bound rheology, a super-adiabatic perovskite layer for the lower-bound rheology, and an azimuthally-averaged viscosity of no more than $10^{26}$ Pa s. Convection in large super-Earths is characterised by large upwellings (even with zero basal heating) and small, time-dependent downwellings, which for large super-Earths merge into broad downwellings. In the context of planetary evolution, if, as is likely, a super-Earth was extremely hot/molten after its formation, it is thus likely that even after billions of years its deep interior is still extremely hot and possibly substantially molten with a "super basal magma ocean" – a larger version of the proposal of (Labrosse et al., 2007), although this depends on presently unknown melt-solid density contrast and solidus.

# 1. Introduction

There is much interest in the possible structure and dynamics of large rocky planets (super-Earths) around other stars, which may exist around as many as 23% of stars (Howard et al., 2010). Initial studies focused on determining the radial structure of such planets (Seager et al., 2007; Sotin et al., 2007; Valencia et al., 2006; Valencia et al., 2007b; Valencia et al., 2007c).

A first-order question is whether super-Earths are likely to experience plate tectonics, and this has so far been approached using simple models. Boundary-layer theory, taking into account the increase in mean density with planet size (Valencia and O'Connell, 2009; Valencia et al., 2007a) predicts that plate tectonics becomes more likely with increasing planet size; recent dynamical calculations and theory support this (Korenaga, 2010a; van Heck and Tackley, 2011). While one study finds the opposite conclusions (O'Neill and Lenardic, 2007), it is argued in (van Heck and Tackley, 2011) that this might be a consequence of working in nondimensional space and not scaling all nondimensional parameters in a consistent manner with planet size.

In general, viscosity and thermal conductivity increase with pressure while thermal expansivity decreases with pressure, all of which result in lower convective vigour in the deep mantle, which several studies have shown to result in large-scale structures in Earth's deep mantle (e.g. (Balachandar et al., 1992; Hansen et al., 1991; Hansen et al., 1993)). The pressure at the core-mantle boundary (CMB) of a ten Earth mass super-Earth is about ten times the pressure at Earth's CMB (Valencia et al., 2009), so if these trends in physical properties continue, a super-Earth's deep mantle would be expected to have a very low effective Rayleigh number and therefore, according to classical ideas, very sluggish or no convection (e.g. (Stamenković et al., 2011)). Stamenković et al. (2012) found using 1-D and 2-D models with an estimated perovskite rheology that in fact the heat transport mode and viscosities depend on initial temperatures and time. The purpose of this study is to investigate the thermal, rheological and convective state with a post-perovskite rheology calculated using density functional theory (DFT).

Dynamical studies to date have generally not taken into account these large changes in viscosity, thermal expansivity and thermal conductivity with pressure. An exception is the study of (van den Berg et al., 2010), who focussed on the influence of variable thermal conductivity on convection in super-Earths that are undergoing stagnant lid convection, finding that large thermal conductivity in the deep mantle results in strong coherent upwellings from the core-mantle boundary. They found that the temperature eventually adopts an adiabatic profile, as is normally found in convective systems. While they also included pressure-dependence of viscosity and thermal expansivity, they did not analyse their effects. Another exception is the study of (Stamenković et al., 2012), who used 1-D parameterized and 2-D convection models to investigate the influence of a strong increase in activation enthalpy with pressure, although they used average scalings for thermal expansivity and thermal conductivity.

Complicating the extrapolation of physical properties to higher pressure is the possible influence of phase transitions. The perovskite (Pv) to post-perovskite (pPv) transition, occurring at a pressure of around 125 GPa, is well-established. An important issue is whether post-perovskite remains stable to ~1400 GPa, which is the pressure at the base of the largest super-Earth's mantle. First-principles calculations by (Umemoto et al., 2006) indicated that $MgSiO_3$ post-perovskite may break down into constituent oxides ($MgO+SiO_2$) at around 1000 GPa. Recent laboratory experiments on the analog material $NaMgF_3$, however, questioned this finding (Grocholski et al., 2010). Thus, for the purposes of this initial study, we assume that post-perovskite is the stable phase to the base of a super-Earth mantle.

The physical properties of post-perovskite are still being evaluated. Its viscosity is highly anisotropic. Ammann et al. (2010b) used density functional theory (DFT) to calculate activation enthapies for the different diffusion directions, finding differences of several orders of magnitude between them. Based on physical arguments they argue that diffusion creep will be controlled by the slowest diffusion direction while dislocation creep could be controlled by that the easiest diffusion direction, concluding that in this case the viscosity of post-perovskite could be 2-3 orders of magnitude lower than that of perovskite at the same pressure and temperature.

While study of (Ammann et al., 2010b) focussed on the pressure range of Earth's mantle, a key question for super-Earths is how the viscosity changes at up to 10 times this pressure. Thus, in this study we have extended these DFT calculations to calculate activation enthalpies for diffusion creep in post-perovskite at pressures of up to 1 TPa. These values are then used in dynamical calculations of mantle convection in super-Earths that also include reasonable physical variations of other parameters and yielding-induced plate tectonics. It has recently been argued that at very high pressures, deformation by interstitial diffusion may become more effective than by vacancy diffusion, possibly even causing in a decrease of viscosity with pressure along an adiabat (Karato, 2011). At present this possibility is not quantified so we will leave it to a future study.

## 2. Rheology

### 2.1 Density Functional Theory calculations

Here we perform density functional theory (DFT) calculations of vacancy diffusion in post-perovskite using the method descibed in (Ammann et al., 2008, 2010a; Ammann et al., 2010b). Full details and accuracy tests are given in Ammann et al. 2010b (main paper and supplementary material); here summarise the main points. The viscosity of post-perovskite, assuming diffusion creep, is controlled by magnesium diffusion and the effective diffusivity, $D_{eff}$, which limits the rate of deformation, can be well approximated by $D_{eff}=5D_{Mg}$. The large anisotropy in diffusion implies a large viscosity spectrum lying somewhere between the two extremal cases of diffusion along x (lower bound) and along y (upper bound). We write the viscosity $\eta$ as follows:

$$\eta = \eta_0 \frac{G^2 k_J T}{N_v} \exp\left(\frac{H}{kT}\right) \qquad (1)$$

Here, $G$ is the characteristic grain size (in the Earth's lower mantle somewhere between 0.1-1 mm), $N_v$ is magnesium-vacancy concentration (in the lower mantle estimated to be around $10^{-7} - 10^{-3}$), $H$ is the migration enthalpy (given in the table below in eV) and $k$ is Boltzmann's constant (8.617343 $10^{-5}$ eV). Grain size and vacancy concentrations are chosen such that they well reproduce lower mantle viscosity profiles (Ammann et al., 2010a) and are assumed to be similar in post-perovskite (Ammann et al., 2010b); a similar vacancy concentration is expected if, as seems probable, vacancies are mostly extrinsic rather than intrinsic. The prefactor is given by $\eta_0 = 1/(5\alpha\, 2/3\, l^2 \nu V_{mol})$ and is listed in Table 1. $\alpha$ is a geometrical factor (40/3 or 16/3 with or without grain boundary sliding respectively, the former was used here), $l$ is the jump distance, $V_{mol}$ is the molecular volume and $\nu$ is the attempt frequency times the exponential of the migration entropy (see (Ammann et al., 2010a; Ammann et al., 2010b) for details). Magnesium diffuses along y via six-jump cycles on the magnesium-silicon sublattice

(Ammann et al., 2010b). For simplicity, we approximate the six different jumps as one single jump with an effective migration enthalpy, $H^{high}$, and attempt frequency (included in $\eta_0$). For magnesium diffusion along x, the fast direction, no such simplification had to be made.

The accuracy of this method was discussed and tested in Ammann et al. (2010b) (supplementary information). Assuming that the difference between LDA (Local Density Approximation) and GGA (Generalized Gradient Approximation) values provides an estimate of the DFT uncertainties, then the energies are correct to within about 10%. The effect of modelled system size was assessed by calculating the migration enthalpies in a 3x1x1 and a larger 3x1x2 supercell, and the values found to differ by less than 5% (in some cases less than 1%) between the two system-sizes (Ammann et al. 2010b supplementary information). System-size effects were investigated in more detail for periclase in Ammann et al. (2012); it was found that migration enthalpies change with system size non-linearly and are already converged to an accuracy of around 10% even in small cells, further supporting the 5% estimated cell-size effect in the present calculations.

**2.2 Analytical fit**

In order to insert $H(p)$ for post-perovskite and perovskite into convection calculations, it is necessary to fit the DFT-calculated values to an analytical expression. After trying theoretical (e.g. elastic strain energy model) and *ad hoc* analytic forms, the following was found to give a good fit:

$$H(p) = E_0 + pV(p) \qquad (2)$$

$$V(p) = V_0 \exp\left(-\frac{p}{p_{decay}}\right) \qquad (3)$$

This was fit to the data in (Ammann et al., 2008) for perovskite and to both the upper-bound and lower-bound data in Table 1 for post-perovskite. For the pressure range of Earth's upper mantle an olivine rheology is assumed (Karato and Wu, 1993). The resulting fit of $H(p)$ is plotted in Figure 1(a) and the relevant parameters are given in Table 2. Clearly, the effective activation volume decreases with pressure and is very low for post-perovskite, dropping over the considered pressure range from about 1.5 to 0.5 cm$^3$/mol for the upper-bound rheology and 1.3 to 0.6 cm$^3$/mol for the lower-bound rheology. This is slightly lower than predicted by (Wagner et al., 2011). Considering total enthalpy, the estimates of Stamenković et al (2011) are within our bounds below 700 GPa, and above 700 GPa are at most 20% larger than our results.

**3. Convection Models**

**3.1 Physical model and numerical method**

*3.1.1 Viscosity*

Diffusion creep is assumed to be the dominant deformation mechanism, and is assumed to follow an Arrhenius law:

$$\eta(T,p) = \eta_0 \exp\left[\frac{H(p)}{RT} - \frac{H(0)}{RT_0}\right] \qquad (4)$$

where $H(p)$ is given by equations (2)-(3), $\eta_0$ is the reference viscosity obtained at zero pressure and reference temperature $T_0$, which is chosen to be 1600 K. As mentioned in section 2.2, an olivine rheology is assumed for the pressure range of Earth's upper mantle, a perovskite rheology is assumed in the perovskite stability field and a post-perovskite rheology is used for higher pressures. For perovskite, $\eta_0$ is calculated so as to give a factor 10 viscosity increase from the upper mantle to the lower mantle, while for post-perovskite $\eta_0$ is calculated so as to give either a factor $10^5$ viscosity increase (for the upper bound pPv rheology) or factor 100 viscosity decrease (for the lower bound pPv rheology) at the pV-pPv transition. The former increase is consistent with Ammann et al. (2010b) while the decrease is somewhat less than the 4-5 orders of magnitude estimated in Ammann et al. (2010b) in order to be conservative regarding the possible weakness of pPv. For this initial study, viscosities in Earth's mantle's pressure range are about 2 orders of magnitude higher than realistic, with upper mantle $\eta_0=10^{21}$ Pa s (resulting values for other phases are listed in Table 2), in order to allow us to resolve convection in a super-Earth with reasonable computational resources.

As discussed in Ammann et al. (2010b), diffusion creep is likely to be controlled by the upper bound rheology; thus we perform a suite of simulations using this. For the sake of comparison, we also perform simulations with the lower bound rheology, which may be relevant to dislocation creep (although for simplicity we here use a linear stress-strain rate relationship for both bounds).

Illustrative resulting viscosity profiles for upper mantle $\eta_0=10^{21}$ Pa s calculated along an adiabat with a 1600 K potential temperature (see next section for relevant physical properties) are plotted in Figure 1(b). Interestingly, while the upper-bound pPv viscosity is extremely high at >$10^{30}$ Pa s, it remains approximately constant with pressure, first increasing by ~1 order of magnitude then decreasing by ~2 orders of magnitude. This is because the viscosity depends on the ratio $H(p)/T_{adiabat}(p)$, and the adiabatic temperature increases in about the same proportion with pressure as $H(p)$. Nevertheless, convection would not take place if the viscosity were this high. The lower-bound pPv viscosity, in contrast, increases by ~7 orders of magnitude along an adiabat from 130 to 1400 GPa. With the assumed absolute viscosities this results in a viscosity of ~$10^{30}$-$10^{31}$ Pa s near the CMB in a 10 $M_E$ super-Earth, which would be enough to stop convection in this region. Even so, this is a much lower increase than the ~15 orders of magnitude predicted by (Stamenković et al., 2011) based on perovskite rheology extrapolated using simple laws. It is possible that a larger viscosity increase than we predict could occur if extrinsic vacancy concentration decreases with pressure, due to changes in element partitioning or growth of a new phase, whereas we assume a constant extrinsic vacancy concentration. However, there is currently little experimental evidence that extrinsic vacancies change significantly with pressure. Nevertheless, one must accept that that vacancy concentrations in super-Earths are uncertain. Larger diffusion-creep viscosities would also occur if grain size becomes larger at higher pressure because the higher temperatures cause more rapid grain growth, as suggested by (Stamenković et al., 2011).

This difference in the pressure scaling of lower- and upper-bound viscosities is due to anisotropic compression of the lattice. Diffusion in the 100 and 001 directions requires dilation in the 101 direction. This is the most compressible direction and so with pressure will become stiffer quicker than in the other two directions. Thus, increasing pressure slows diffusion in the 100 and 001 directions to a greater extent than diffusion in the 010 direction (the slowest one). Note that the "lower-bound" viscosity assumed here is 2-3 orders of magnitude higher (compared to Pv) than predicted in Ammann et al. (2010b) in order to be conservative; if we plotted the latter then it would always be substantially lower than the upper-bound viscosity. If the lower bound is relevant to dislocation creep, then the power-law rheology associated with dislocation creep would likely further reduce the viscosity increase, because Christensen (1984) showed that power-law rheology can be approximated by Newtonian rheology with its activation enthalpy reduced by factor n, the power-law index.

Plastic yielding occurs at high stresses, giving a first-order approximation of plate tectonics (Moresi and Solomatov, 1998; Tackley, 2000; Trompert and Hansen, 1998). The pressure-dependent yield stress is given by a combination of brittle and ductile processes as:

$$\sigma_Y = \min\left(c_f p \, , \, \sigma_{duct} + \sigma'_{duct} p\right) \quad (5)$$

where $c_f$ is the Byerlee's law friction coefficient, $\sigma_{duct}$ is the ductile yield stress and $\sigma'_{duct}$ is the pressure gradient of ductile yield stress, which prevents yielding in the deep mantle. Values are given in Table 4. For the smallest planet case, $\sigma_{duct}$ had to be reduced by a factor of 2 in order to avoid the stagnant lid mode of convection.

The effective viscosity, combining diffusion creep and plastic yielding, is given by:

$$\eta_{eff} = \left(\eta_{diff}^{-1} + \frac{2\dot{e}}{\sigma_Y}\right)^{-1} \quad (6)$$

where $\dot{e}$ is the second invariant of the strain-rate tensor.
Finally, the viscosity is truncated between $10^{19}$ and $10^{40}$ Pa s (using simple min and max functions) in order to facilitate numerical solution.

*3.1.2 Density*

A Birch-Murnaghan third order equation of state is used to relate density to pressure and calculate bulk modulus K. For simplification, rather than consider all mineral phases present the mantle is divided into "upper mantle", "transition zone", "perovskite" and "post-perovskite" mineralogies, with different parameters ($K_0$, K' and $\varrho_0$) for each, which are chosen to match the widely-used Earth model PREM (Dziewonski and Anderson, 1981). Obtained values are listed in Table 3 and the resulting density profile along the reference adiabat is given in Figure 2b.

*3.1.3 Thermal expansivity*

The thermal expansivity depends on pressure and is given by:

$$\alpha = \frac{\rho \gamma C_p}{K} \quad (7)$$

where the pressure-dependent density $\varrho$ and bulk modulus K are calculated as described above, the specific heat capacity $C_p$ is assumed constant, and Gruneisen parameter $\Upsilon$ is given by:

$$\gamma = \gamma_0 \left(\frac{\rho_0}{\rho}\right). \quad (8)$$

This leads to the profile in Figure 2(c). The effect of Gruneisen parameter decreasing with density (pressure) is to give a stronger decrease of thermal expansivity with pressure, and therefore decrease the adiabatic temperature gradient, making the interior cooler (hence more viscous) than would be predicted with a constant Gruneisen parameter. Thermal expansivity decreases to about $0.9 \times 10^{-5}$ K$^{-1}$ at Earth's CMB pressure, consistent with mineral physics constraints (Chopelas and Boehler, 1992; Katsura et al., 2010; Komabayashi et al., 2008; Mosenfelder et al., 2009), and to about $1.5 \times 10^{-6}$ K$^{-1}$ at 1400 GPa, which appears to be consistent with that estimated by (Stamenković et al., 2011) using a completely different approach.

This leads to the adiabatic temperature profile given in Figure 2(a), in which a 1600 K surface temperature increases to ~2350 K at Earth's CMB pressure and ~3630 K at a 10 $M_E$ super-Earth's CMB. This corresponds to mean dissipation numbers (given by $\log(T_{cmb}/T_{surf})$) of 0.38 and 0.82, respectively. The dissipation number gives an indication of the importance of viscous dissipation and adiabatic heating in the global energy balance (Backus, 1975; Hewitt et al., 1975; Jarvis and McKenzie, 1980). Thus, despite the much higher pressure, these terms are only ~twice as important as they are in Earth.

*3.1.4 Thermal conductivity*

The thermal conductivity is $k = \rho C_p \kappa$, where the thermal diffusivity $\kappa$ is given by:

$$\kappa = \kappa_0 \left(\frac{\rho_0}{\rho}\right)^{-1} \tag{9}$$

This gives a moderate increase of $k$ with pressure that is consistent with findings from a recent mineral physics study (de Koker, 2010) although other studies estimate higher values (Goncharov et al., 2009; Hofmeister, 2008). The surface value of $k$ is 3.0 W/K/m, and over Earth's mantle's pressure range it increases to 7.2 W/K/m. This treatment does not include the radiative component. It is plausible that the radiative component might be important at the high temperatures obtained deep in super-Earths, but for simplicity it will not be considered in this initial study.

*3.1.5 Constant properties and boundary conditions*

Internal heating rate per unit mass is assumed constant, as is heat capacity $C_p$. Acceleration due to gravity $g$ is also assumed to be constant but varies with planet size. In Earth's mantle this is a good approximation because $g$ varies by only ~8% according to PREM (Dziewonski and Anderson, 1981). In a 10 $M_E$ super-Earth $g$ may change by a somewhat larger factor, but as this is smaller than the uncertainties in most other physical parameters, it is reasonable to ignore it and assume a constant value. This constant value is set so as to give the correct pressure at the CMB for a planet of a given size.

The surface is assumed to be isothermal with a temperature of 300 K, while the CMB is also assumed to be isothermal, with a temperature that is adjusted each time step to give zero net heat flow. Zero net CMB heat flow is an idealisation made for this initial study because we have no idea what the CMB temperature is in super-Earths: predicting this requires detailed modelling of their evolution from a hot initial state, which has so far been done only for isoviscous super-Earths (Kite et al., 2009; Papuc and Davies, 2008) – we plan in future to do this for super-Earths with pressure-dependent viscosity. In Earth, it is generally thought that heat from the core provides only a small fraction of the global heat budget (Davies and

Richards, 1992; Jaupart et al., 2007; Schubert et al., 2000), therefore this is a reasonable first-order approximation to ignore it. Both boundaries are mechanically free-slip.

*3.1.6 Planetary dimensions*

It is assumed that the relative masses of the mantle and core, as well as their compositions, are the same as those of the Earth. For a given planetary mass, the planetary radius, CMB radius and the average $g$ for the mantle are calculated using a Matlab script that calculates a 1-D profile of the planet by integrating with pressure the equations given above, and iterating on planet size until the mass and size match. The results are in Table 5. These dimensions and the mean $g$ are then used as input parameters for the convection calculations. These numbers are slightly different from those calculated by Valencia et al. (2006) because they assumed a pure iron core whereas here the parameters are tuned to match PREM (Dziewonski and Anderson, 1981). These differences are typically of order ~few %; up to 9% for pressure at the CMB, which reflects the general uncertainty in calculating high-pressure properties and will not make a significant difference to the convection results presented here.

*3.1.7 Phase transitions*

Phase transitions are included, as in our previous studies for Earth (e.g. (Nakagawa and Tackley, 2005b, 2011; Tackley and Xie, 2003)), but we here set all Clapeyron slopes to zero in order to reduce the number of complexities influencing the convection; phase transitions thus influence only the radial density profile and the viscosity. The reader is referred to publications referenced above for full details on the assumptions and implementation. In brief: two sets of phase transitions are included: those for olivine and those for pyroxene-garnet, and it assumed that the mantle is 60% olivine and 40% pyroxene-garnet. Table 6 gives the included phase transitions. For super-Earths, depths are scaled inversely with mean gravitational acceleration listed in Table 5.

*3.1.8 Solution method*

The physical model is solved using the numerical code StagYY (Tackley, 2008) in a two-dimensional spherical annulus (Hernlund and Tackley, 2008). StagYY uses a finite-volume discretization of the governing compressible anelastic equations, and either a multigrid solver or a direct solver to obtain a solution. For the present calculations the direct solver UMFPACK (Davis, 2004) is used, and is accessed using the PETSc toolkit (Balay et al., 2012; http://www.mcs.anl.gov/petsc). For this study, dimensional units are used throughout because most nondimensional parameters involve the mantle depth $D$ as the length scale, so changing planet size would result in almost all nondimensional parameters changing, but only a few dimensional parameters changing. For this initial study, calculations are run to a statistically steady-state, at which quantities such as heat flux, mean temperature and rms. velocity fluctuate about a mean value that does not have a secular change. This can take billions of years of simulated time. The numerical resolution depends on planet size: higher resolution is needed for larger planets. When a statistically steady-state is reached the solution is continued at twice the resolution in both directions in order to check that it does not change significantly. The highest resolution used was 128 points in radius by 1024 points in azimuth, with radial grid spacing decreased by a factor of 2-3 towards the surface in order to adequately resolve the lithosphere.

**3.2 Results**

*3.2.1 Upper bound rheology*

Figure 3 shows temperature and viscosity fields for all five planet sizes and the upper bound rheology. The existence of downwellings with a low temperature and high viscosity in all planet sizes indicate a mobile-lid tectonic mode, i.e. the lithosphere can "subduct". If instead the stagnant lid mode were present then downwellings would have a very small temperature contrast of order the rheological temperature scale (Solomatov, 1995) and would therefore be barely visible or invisible in these plots. Inspection of the viscosity fields indicates a thin high-viscosity lithosphere; thinner on larger planets as expected from basic scalings (e.g. Valencia et al., 2007a, van Heck and Tackley, 2011). This mobile lid tectonic mode is not exactly Earth-like plate tectonics, however: "slabs" seem to aggregate into large-scale highly viscous downwellings due to the slow, large-scale flow enforced by the high-viscosity post-perovskite layer.

Sluggish convection exists in the deep mantles of all planet sizes because the post-perovskite layers are hot and therefore the viscosity is lower than that predicted along a reference adiabat: in the range $10^{22}$-$10^{28}$ Pa s. Several large-scale hot plumes emanate from a particularly hot region near the CMB, despite the lack of heat from the core. These large-scale plumes feed narrow plumes in the perovskite layer (visible particularly in the viscosity field).

*3.2.2 Lower bound rheology*

Figure 4 shows temperature and viscosity fields for all five planet sizes and the lower bound rheology. Again, the existence of downwellings with low temperatures and high viscosities in all planet sizes indicates that the lithosphere is in a mobile lid, and not stagnant lid, tectonic mode. The lithosphere is generally strong but with some weak zones, and one or more subducted "slabs" are visible. In the smaller planets (1-3 $M_E$) the slabs eventually sink to the deep mantle, but in larger planets they stop before sinking that far, although they do sink into the post-perovskite layer. In larger planets "slabs" aggregate into large, broad downwellings, as in the upper-bound cases (Figure 3).

The thermal structure in the deep mantle changes dramatically with planet size. In small (1-3 $M_E$) planets, subducted material sinks to the region above the CMB resulting in a cool, subadiabatic deep mantle. In the viscosity plot the strong lithosphere, relatively weak upper mantle, and the cool, higher-viscosity region in the deep mantle are clearly visible. Large (5-10 $M_E$) planets display quite different behaviour. The deep mantle is very hot - hot enough that material can flow due to the temperature-dependence of viscosity. Several large plumes form from this hot material (despite zero net heat flow across the CMB) and rise to the base of the lithosphere. The deep-mantle viscosity is high, but not as high as it would be along an adiabatic temperature profile. Slabs pool above this deep high-viscosity region.

Thus, all size planets have a relatively high viscosity in the deep mantle but for different reasons. In small planets it is caused by cold slabs pooling, whereas in large planets it is caused by the intrinsic increase of viscosity with pressure. The deep mantle temperature is quite different: subadiabatic in small planets but superadiabatic in large planets.

*3.2.3 Radial profiles*

These findings are emphasised in profiles of azimuthally-averaged temperature and viscosity (Figure 5). For upper-bound cases, the temperature in the post-perovskite region is strongly elevated above the 1800 K adiabat - by around 1000 K near the base of their mantles. As a result, the azimuthally-averaged viscosity is much lower than that predicted in Figure 1, being in the range $10^{25}$-$10^{26}$ Pa s in most of the pPv layer, but decreasing to ~$10^{24}$ Pa s near the base of the post-perovskite layer. The pPv layer is around 2-3 orders of magnitude more viscous than the perovskite layer.

The lower-bound cases display distinctly different radial profiles, but also with higher than adiabatic temperatures. Up to ~300 GPa, temperature profiles are either adiabatic (5-10 $M_E$) or subadiabatic (1-3 $M_E$). Above 300 GPa, the temperature profile is superadiabatic, following approximately the same profile for all planet sizes 5-10 $M_E$, except towards the base of the largest planets (5-10 $M_E$), where the temperature increases more rapidly. Viscosity profiles show an increase of viscosity with pressure at lower pressures, but once the viscosity has reached ~$10^{25}$-$10^{26}$ Pa s it does not increase any further.

Thus, there is a self-regulation of viscosity due to the fact that in equilibrium, each part of the mantle must be losing its radiogenically-produced heat. If, at some time, the viscosity is too low (temperature is too high) then convection will be more vigorous and cooling will take place; on the other hand if viscosity is too high (temperature too low) for the heat to be advectively lost, then the mantle heats up, reducing the viscosity until material can flow. The former scenario is more relevant to planetary evolution, as discussed below.

**4. Discussion and conclusions**

The calculations presented here indicate the dominant influence of rheology on the dynamics of super-Earth mantles, focussing on potentially habitable super-Earths that have the same surface conditions as Earth. The rheology used, in particular activation enthaply as a function of pressure $H(p)$, is based on published DFT calculations for perovskite (Ammann et al., 2008) and new calculations for post-perovskite (extending (Ammann et al., 2010b)). We consider rheologies based on both the slowest diffusion direction (upper bound), which probably dominates diffusion creep, and the fastest diffusion direction (lower bound), which has been argued to be dominant in dislocation creep (Ammann et al., 2010b). Our convection calculations show that the deep mantles of large super-Earths (7-10 $M_E$) convect, even if their viscosity calculated assuming an adiabatic temperature profile is so high that they are expected not to. This is due to internal heating coupled to the feedback between internal heating, viscosity and temperature: the deep mantle simply adjusts its temperature such that its viscosity is the value needed for it to lose its radiogenic heat. A similar self-regulation was proposed by Tozer (1972) but for the average mantle viscosity; here it applies locally rather than globally. With the upper-bound viscosity the entire post-perovskite layer is hot, whereas with the lower-bound viscosity a super-adiabatic temperature profile results. The maximum azimuthally-averaged viscosity in both cases is in the range $10^{25}$-$10^{26}$ Pa s for the parameters assumed here.

Analytical theory by (Fowler, 1993) predicted that convection in the limit of strongly pressure- and temperature- dependent viscosity tends to adopt an isoviscous profile. While previous numerical convection studies for Earth have not obtained such a profile, we here find that this theory does indeed apply to very large planets with internal heating. Smaller planets do not have a sufficiently large viscosity increase along an adiabat for this asymptotic limit to be reached. Earlier super-Earth calculations with depth-dependent viscosity by (van den Berg et al., 2010) also did not reach this asymptotic state, instead finding an adiabatic temperature profile after several billion years of evolution. Recent two-dimensional calculations (Stamenković, et al., 2012) did obtain a super-adiabatic temperature profile in the case of a strong increase of activation enthalpy with pressure, similar to our results presented here, furthermore finding that this state could take a long time to reach depending on the initial temperature. Recent parameterised studies that use mixing-length theory rather than the usual boundary layer theory to model heat transport in the mantle (Tachinami et al., 2011; Wagner et al., 2011) also found such a self-regulation effect; their agreement with the full convection models presented here suggests that mixing length theory is a suitable parameterisation to use in this situation. Thus, for a pPv rheology that is as realistic, as far as we can calculate and consistent with some previous studies, a deep stagnant layer cannot form.

The present calculations assume a viscosity that is approximately two orders of magnitude higher than Earth's (at a particular temperature and pressure), in order to properly resolve the dynamics in the largest planets. This is unlikely to change the basic physics and the trends observed here but will make some quantitative differences. If the viscosity were two orders of magnitude lower at every temperature and pressure, a super-Earth's deep mantle would not have to heat up as much to lose its radiogenic heat. The viscosity would still adjust to $\sim 10^{25}$-$10^{26}$ Pa s, but the temperature would be less super-adiabatic than predicted here.

There is, on the other hand, considerable uncertainty in the absolute viscosity of post-perovskite (of which super-Earth mantles are mainly comprised) due to unknown grain size and vacancy concentration, and also to the highly anisotropic viscosity. Here we assume that post-perovskite has a five orders of magnitude higher viscosity than perovskite at Earth deep mantle convections if the upper bound rheology dominates, or a two orders of magnitude lower viscosity than perovskite if the lower bound rheology dominates. If it were assumed that post-perovskite is more viscous, then super-Earth mantles would be hotter; conversely if we assumed that post-perovskite is less viscous then super-Earth mantles would be less hot. If dislocation creep is important, then this would tend to reduce the viscosity because, to first order, power-law rheology can be approximated by linear rheology with an activation enthalpy reduced by factor n, the power-law exponent (Christensen, 1984). Another possibility is that at high pressures the creep mechanism changes to intersticial diffusion and that this could cause a viscosity reduction with pressure (Karato, 2011), in which case a super-mantle could be adiabatic. At present there is no quantitative data available to assess this.

Here we focus on calculations that are in thermal equilibrium, with a surface heat flux that matches the heat input by radiogenic heating, and thus with no secular change in temperature. We first note that sufficient internal heating (a high Urey ratio) is necessary to obtain the self-regulation of viscosity observed here. Secondly, in reality planets cool from a hot, and possibly molten, initial state. Previously it has been found that statistically-steady-state calculations give a good approximation to the dynamics of cooling-planet calculations (e.g. the Nu-Ra relationship is the same), except that cooling appears as an additional effective heat source (e.g. Honda and Iwase, 1996). A prediction of the influence of cooling from a hot initial state can be found in the models of Tachinami et al (2011), who use mixing-length theory to predict 1-D profiles of a super-Earth as a function of time. From a hot initial state they find that the mantle of a 5 $M_E$ super-Earth fairly rapidly (within 1-2 Gyr) adopts a super-adiabatic profile similar to that found in our lower-bound cases, then changes quite slowly over subsequent billions of years. The initial adjustment is rapid because convective heat loss is rapid when the mantle is hot and hence low viscosity. However, subsequent 1-D models using a different parameterization (Stamenković et al., 2012) found a somewhat longer adjustment time from a hot initial state, so the question needs to be resolved in the future using full convection models. Another effect that would influence cooling-planet calculations is heat conducted from the metallic core, which as a result of core formation is expected to start super-heated relative to the mantle (Stevenson, 1990).

It has been proposed that at early times Earth's mantle had both a shallow magma ocean and a deep, basal magma ocean (BMO) (Labrosse et al., 2007). Because heat loss from the BMO is controlled by the high-viscosity mantle, the BMO could persist for up to billions of years, with remnants of it left today as partially-molten ultra-low velocity zones (ULVZs) observed seismically above the CMB (Lay et al., 2004). Larger planets evolve more slowly with time, as indicated by parameterised convection models (Kite et al., 2009; Papuc and Davies, 2008). Because these parameterised models do not account for high viscosity in the deep mantle, they are likely to overestimate cooling of the deep mantle, which would be slowed by the high deep-mantle viscosity and resulting high temperatures found here. Therefore, if super-Earths started off molten, it is plausible that their deep mantles could retain a super-BMO for many billions of years.

Although the present study makes no attempt to systematically study the question of plate tectonics on super-Earths and we have limited our study to potentially habitable Earth-like super-Earths, it is notable that mobile-lid behaviour occurs for all planet sizes studied here while using the same lithospheric yielding parameters (except for the smallest planet where yield stress had to be decreased) and surface temperature. This is consistent with predictions from parameterised (Valencia and O'Connell, 2009; Valencia et al., 2007a) and simplified numerical (Korenaga, 2010a; van Heck and Tackley, 2011) studies. However, it is important to note that many factors other than planet size influence plate tectonics, including surface temperature (Lenardic et al., 2008; Landuyt et al., 2009; Foley et al., 2012), the presence of liquid water (e.g. Regenauer-Lieb et al., 2001), and internal heating rate (O'Neill et al., 2007). Furthermore, the simple plastic yielding assumed here does not account for history-dependence of rheology, which may be important for plate tectonics (e.g. Landuyt et al., 2009; Foley et al., 2012). Clearly, much additional work is needed in order to obtain a systematic understanding of the likelihood of plate tectonics on super-Earths.

Possible breakdown of post-perovskite into constituent oxides as predicted by (Umemoto et al., 2006) but questioned by (Grocholski et al., 2010) is also not treated here. Currently we have no estimates of the activation energies in these minerals, so the possible influence on dynamics is impossible to predict. Metallization of the major minerals is another possibility, which would not necessarily change activation energies, but would certainly change whether radiative heat transport is important or not; moreover, heat could then be transported via electrons. More efficient conductive heat transport would reduce the need for advective heat transport, allowing viscosities to increase somewhat. There seems to be an indication of a 'last' phase transition within oxide mantles. Mashimo et al. (2006) showed that the transition to a virtually incompressible oxide $Gd_3Ga_5O_{12}$ happened after 1.20 Mbars and suggested that "similar quasi incompressible oxide phases composed of Si, Fe, Mg, an other elements with relatively high natural abundances, rather than Gd and Ga, might exist in the deep mantles of large extrasolar rocky planets".

The models presented here do not account for chemical differentiation, which is thought to influence lithospheric dynamics by forming buoyant crust and water-depleted stiff lithosphere (e.g. (Korenaga, 2010b)), and to influence deep-mantle dynamics by the accumulation of dense, iron rich material above the CMB (e.g. (Nakagawa and Tackley, 2005a, 2010)). They also do not include other effects of melting such as magmatic heat transport, which can be an important heat loss mechanism in early planetary evolution (Nakagawa and Tackley, 2012).

In summary, the models presented here can be regarded as a preliminary assessment of the influence of a realistic activation enthaply profile on mantle dynamics in super-Earths. The influences of rheological uncertainties, cooling from a hot initial state, widespread melting, and differentiation, need to be assessed in future studies in the context of thermo-chemical evolution of super-Earths.

**Acknowledgments.** We thank Dave May for helping interface StagYY to PETSc, David Bercovici and Vlada Stamenković for helpful comments that improved the manuscript, and Oded Aharonson for his editorial work.

**Figure Captions**

**Figure 1.** (Top): Activation enthapy as a function of pressure for perovskite and post-perovskite. The points are from the Density Functional Theory of Ammann et al (2009) for perovskite and Ammann et al. (2010) and the present study for post-perovskite. The lines are the analytical fits. (Middle): Viscosity profile along the 1600 K adiabat for the upper-bound rheology. (Bottom) Viscosity profile along the 1600 K adiabat for the lower-bound rheology.

**Figure 2**. Pressure-dependence of density, thermal expansivity and thermal conductivity, and the resulting 1600 K adiabat, as labelled.

**Figure 3**. Representative viscosity (left column) and temperature (right column) fields for the upper-bound rheology and all five planet sizes. For clearer visualization the viscosity is truncated at $10^{28}$ Pa s, which is lower than the $10^{40}$ Pa s truncation used in the numerical calculations.

**Figure 4**. Representative viscosity (left column) and temperature (right column) fields for the lower-bound rheology and all five planet sizes. For clearer visualization the viscosity is truncated at $10^{28}$ Pa s, which is lower than the $10^{40}$ Pa s truncation used in the numerical calculations.

**Figure 5**. Profiles of azimuthally-averaged temperature and viscosity for five planet sizes, as labelled. The arithmetic average is used for temperature and the geometric average for viscosity. In the temperature, the 1800 K adiabat is also plotted.

**Tables**

| Pressure (GPa) | $\eta_0^{low}$ | $H^{low}$ (eV) | $\eta_0^{high}$ | $H^{high}$ (eV) |
|---|---|---|---|---|
| 131.6 | 2185x10$^{30}$ | 3.4 | 40x10$^{30}$ | 10.3 |
| 505.4 | 1245x10$^{30}$ | 7.1 | 95x10$^{30}$ | 14.0 |
| 1016.1 | 945x10$^{30}$ | 9.6 | 180x10$^{30}$ | 15.2 |

**Table 1**: Quantities required for the viscosity of MgSiO$_3$ post-perovskite.

| Mineralogy | $E_0$ (kJ/mol) | $V_0$ (cm$^3$/mol) | $p_{decay}$ (GPa) | $\eta_0$ (Pa s) |
|---|---|---|---|---|
| "upper mantle" | 300 | 5.00 | $\infty$ | 10$^{21}$ |
| perovskite | 370 | 3.65 | 200 | 3.00x10$^{23}$ |
| post-perovskite lower bound | 162 | 1.40 | 1610 | 1.90x10$^{21}$ |
| post-perovskite upper bound | 780 | 1.70 | 1100 | 1.05x10$^{34}$ |

**Table 2**. Parameters for H(p) analytical fit (equations (2)-(4)).

| Mineralogy | $K_0$ (GPa) | K' | $\gamma_0$ |
|---|---|---|---|
| "upper mantle" | 163 | 4.0 | 1.3 |
| "transition zone" | 85 | 4.0 | 0.85 |
| "perovskite" | 210 | 3.9 | 1.3 |
| "post-perovskite" | 210 | 3.9 | 1.3 |

**Table 3**. Fit of Birch-Murnaghan equation of state parameters for four mantle mineralogies, plus assumed zero-pressure Gruneisen parameter.

| Property | Symbol | Value | Units |
|---|---|---|---|
| Conductivity: surface | $k_0$ | 3.0 | W/m-K |
| Heat Capacity | Cp | 1200 | J/kg-K |
| Surface temperature | $T_{surf}$ | 300 | K |
| Internal heating rate | H | 5.2x10$^{-12}$ | W/kg |
| Friction coefficient | $c_f$ | 0.1 | - |
| Yield stress (p=0) | $\sigma_{duct}$ | 50 (1 $M_E$) <br> 100 (3-10 $M_E$) | MPa |
| Yield stress pressure gradient | $\sigma'_{duct}$ | 0.01 | - |

**Table 4**. Values of various constant properties.

| M/M$_E$ | R (km) | R$_{cmb}$ (km) | <g>$_{mantle}$ (m/s$^2$) | P$_{cmb}$ (GPa) |
|---|---|---|---|---|
| 3 | 8578.0 | 4602.1 | 16.996 | 388.4 |
| 5 | 9823.6 | 5186.8 | 22.062 | 657.9 |
| 7 | 10703 | 5597.6 | 26.382 | 944.5 |
| 10 | 11683 | 6063.5 | 32.076 | 1400 |

**Table 5**. Calculated dimensions and mean gravitational acceleration of the mantle as a function of planet mass.

| Depth (km) | Temperature (K) | Δρ (kg/m$^3$) |
|---|---|---|
| *Olivine-Spinel-Perovskite-Postperovskite* | | |
| 410 | 1600 | 280 |
| 660 | 1900 | 400 |
| 2740 | 2650 | 60 |
| *Pyroxene-Garnet-Perovskite-Postperovskite* | | |
| 60 | 0 | 350 |
| 400 | 1600 | 100 |
| 720 | 1900 | 500 |
| 2700 | 2650 | 60 |

**Table 6**. Phase transitions and properties. All Clapeyron slopes are set to 0

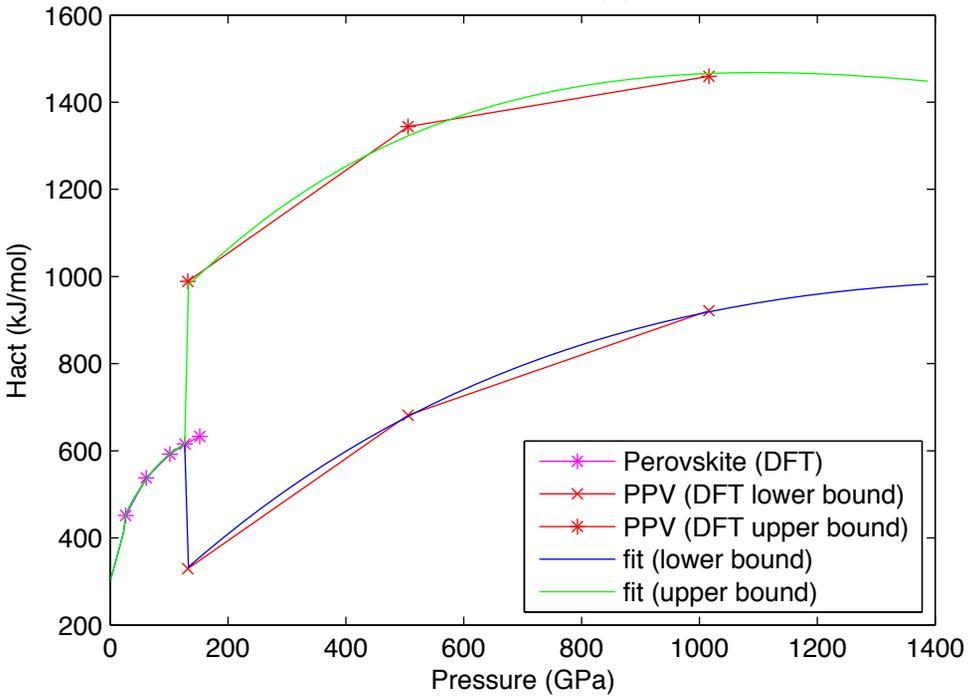

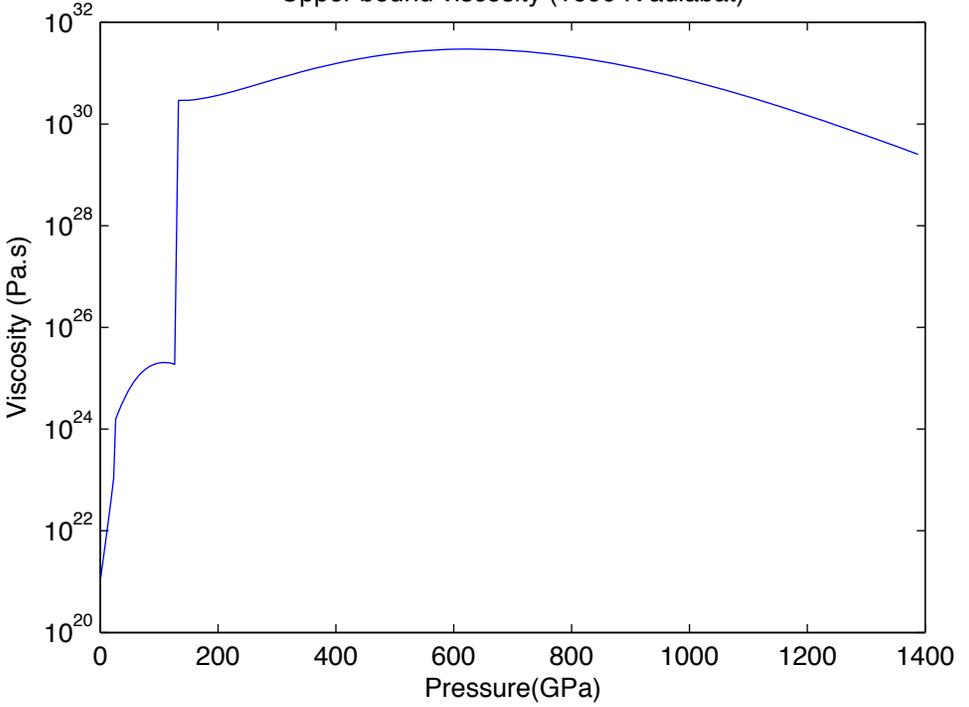

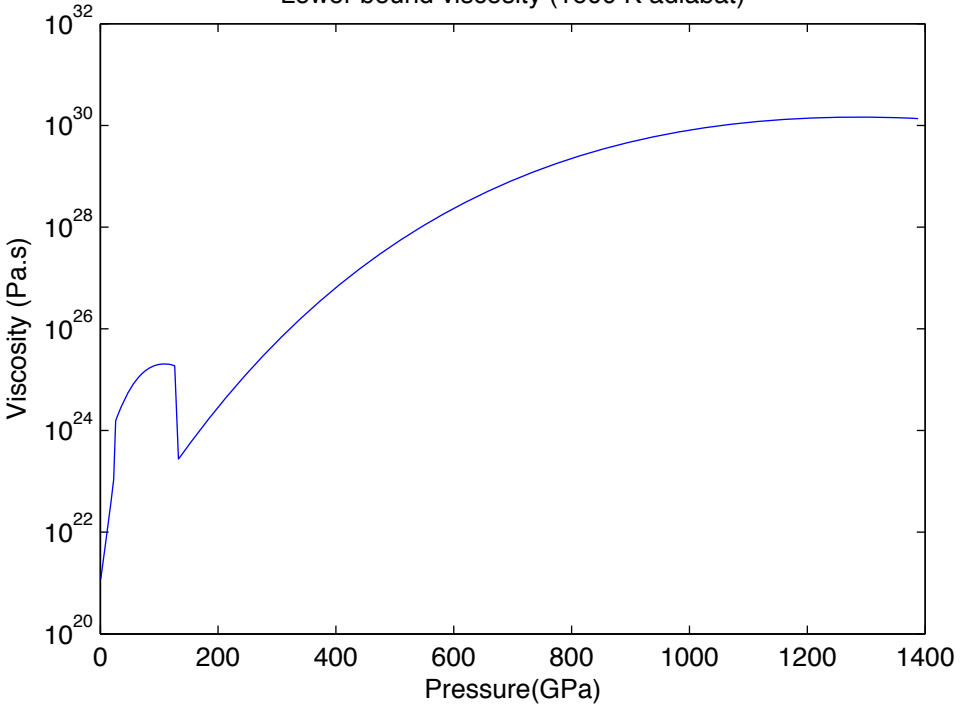

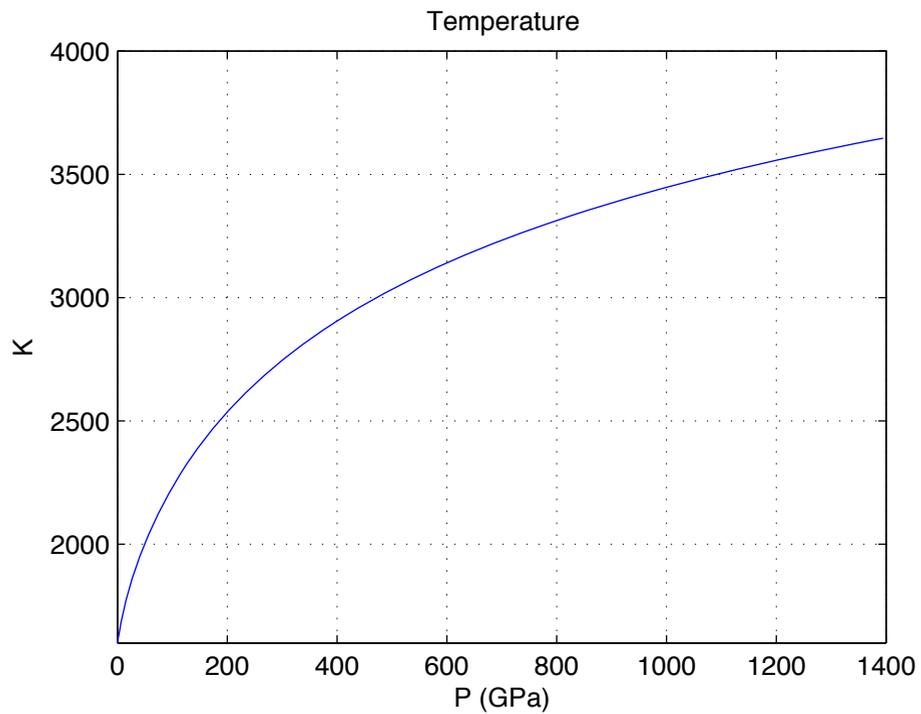
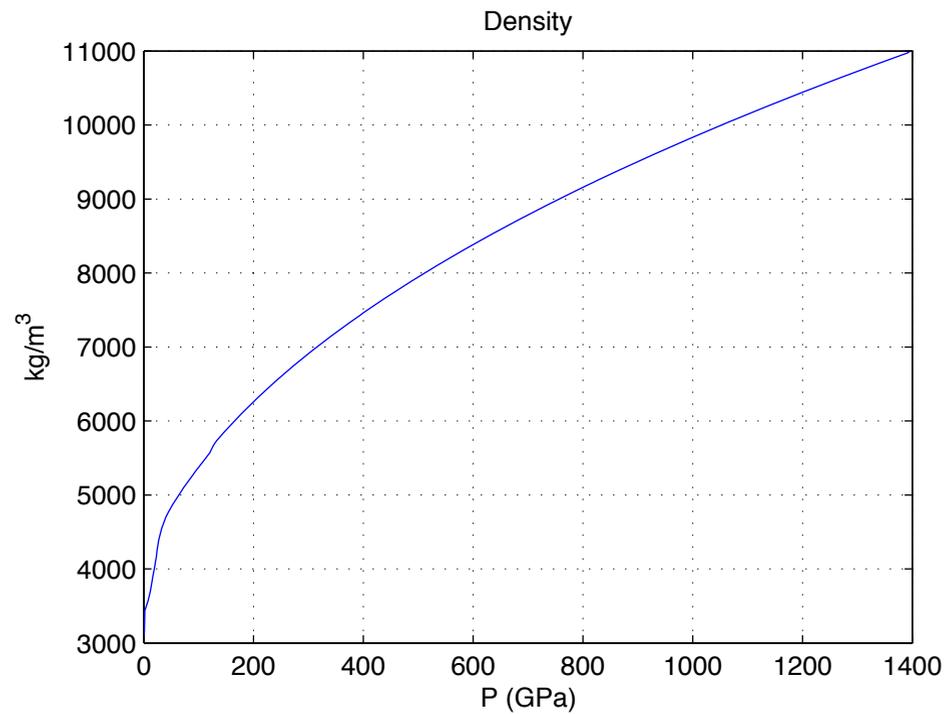
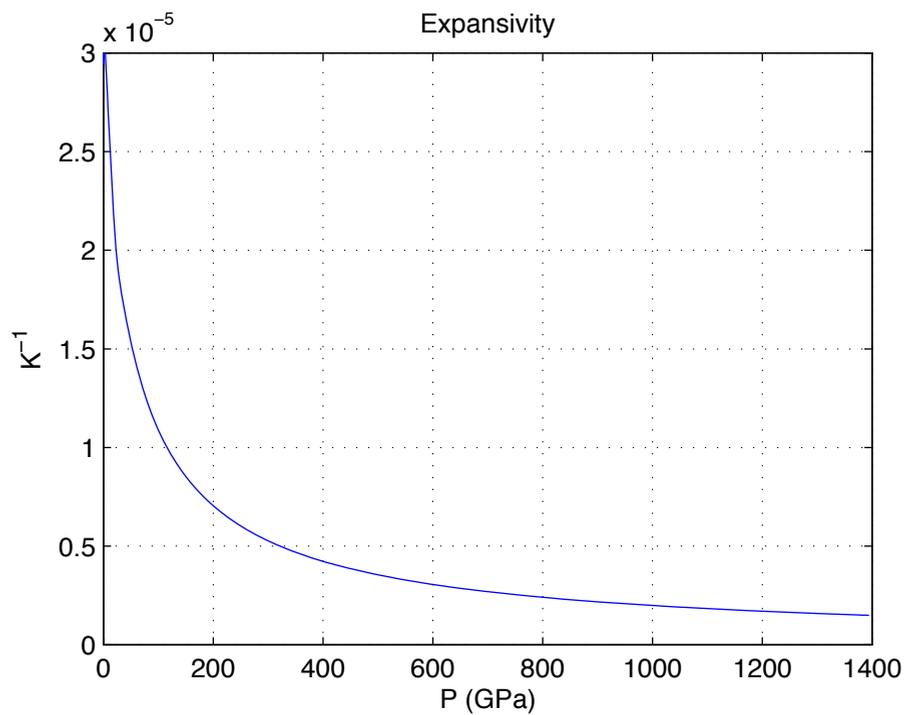
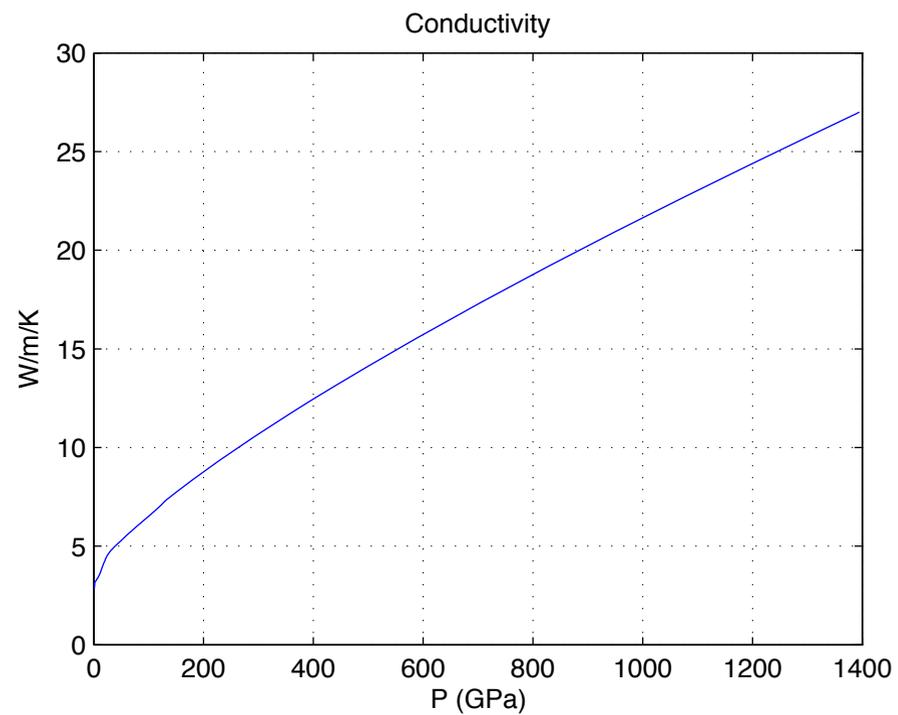

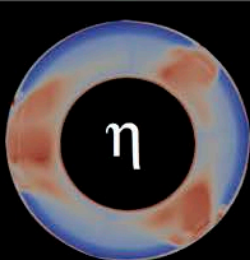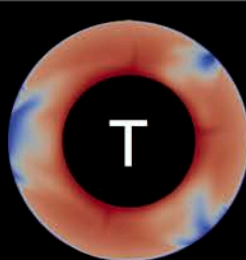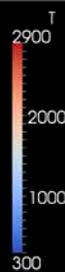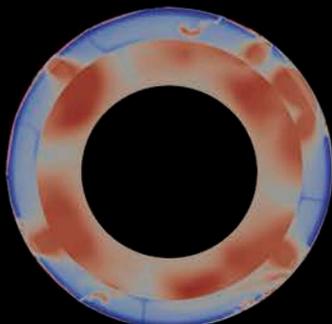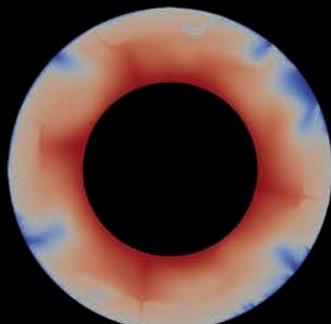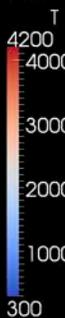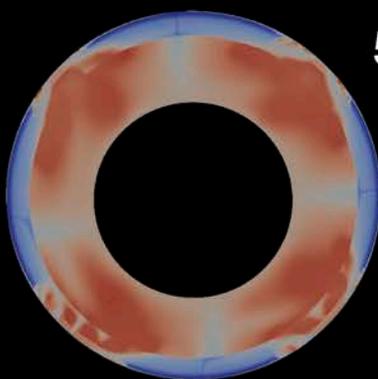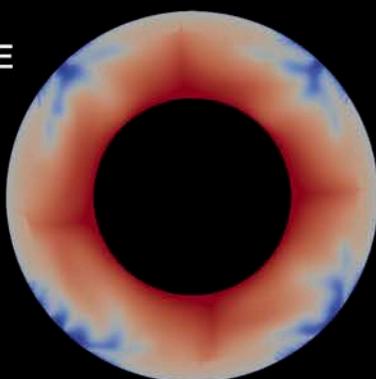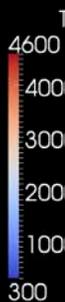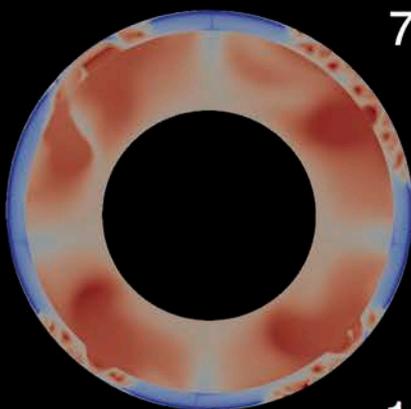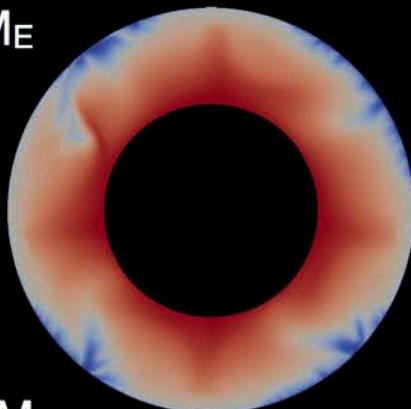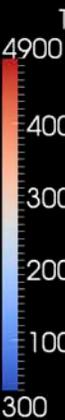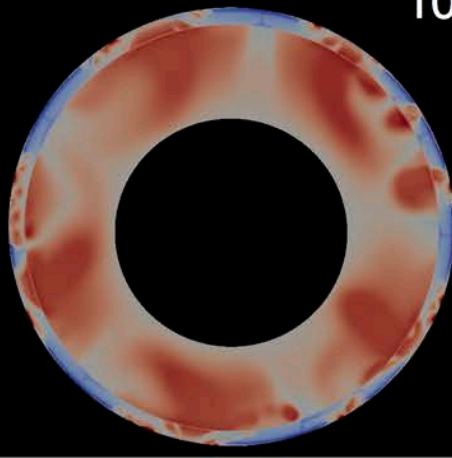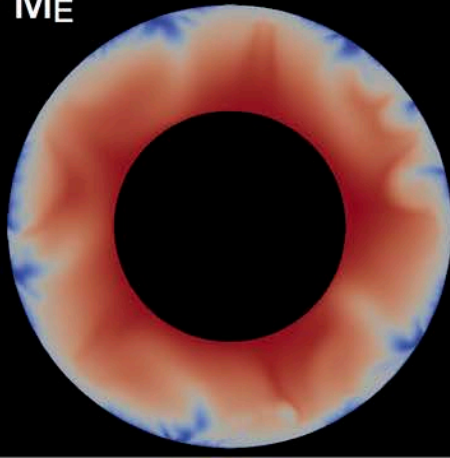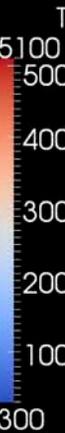

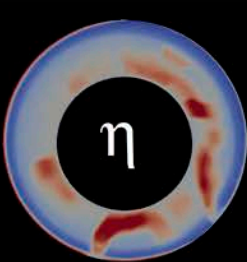 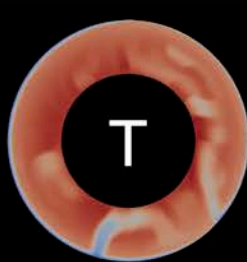 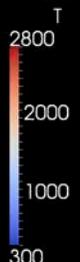

1 $M_E$

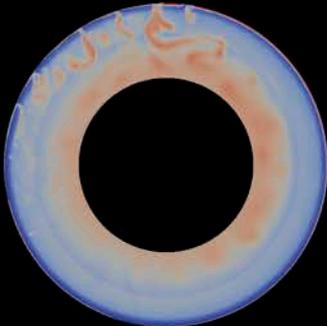 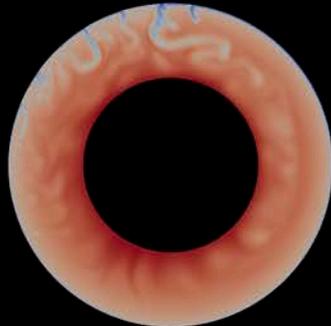 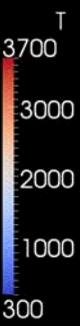

3 $M_E$

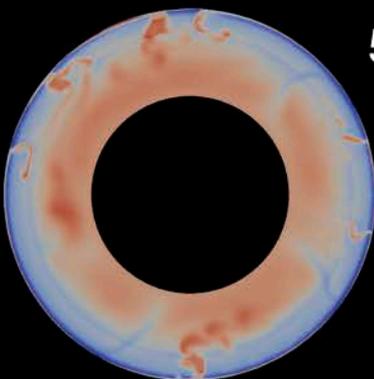 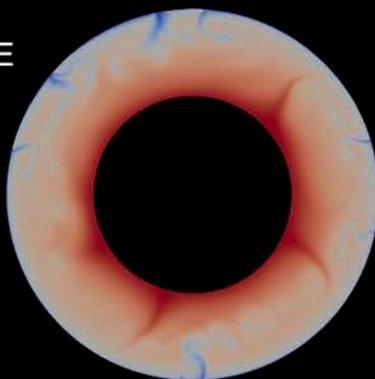 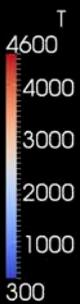

5 $M_E$

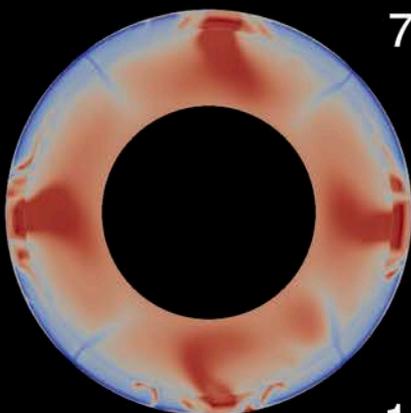 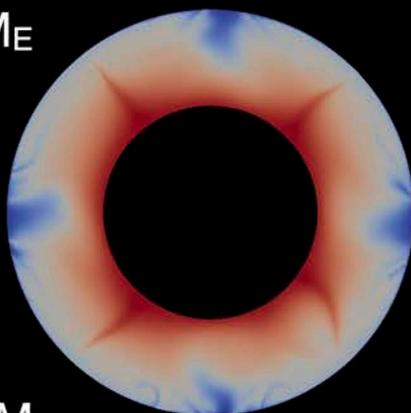 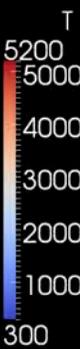

7 $M_E$

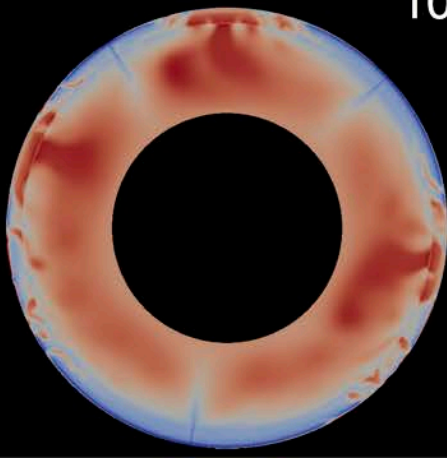 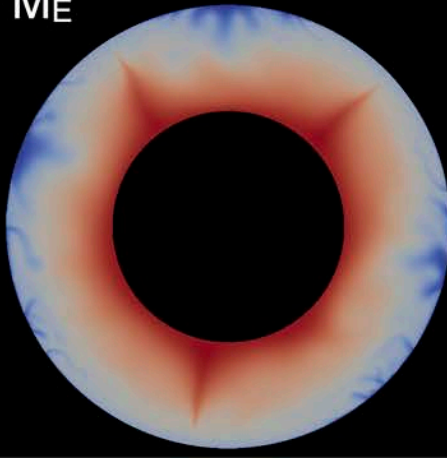 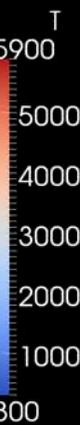

10 $M_E$

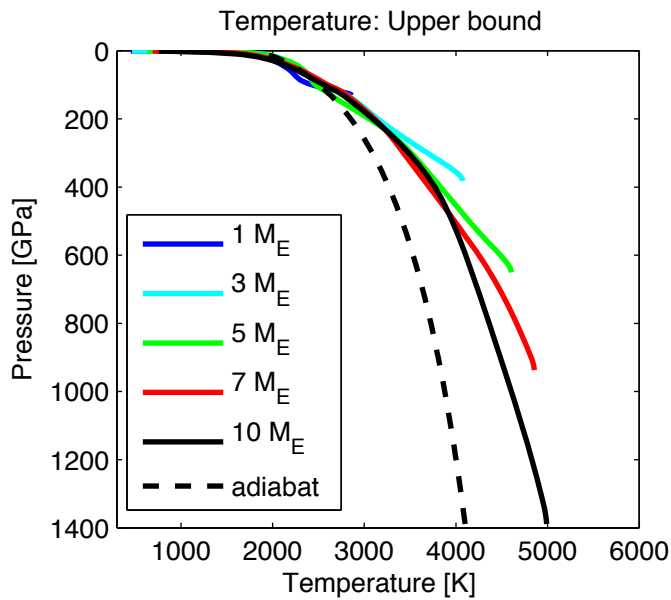
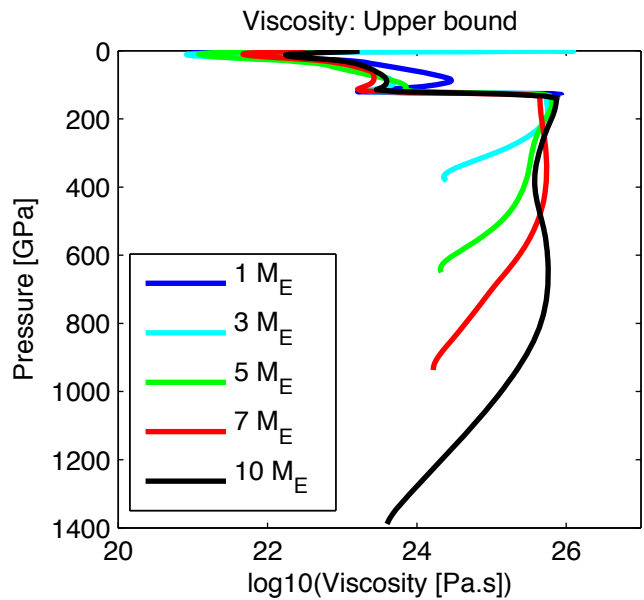
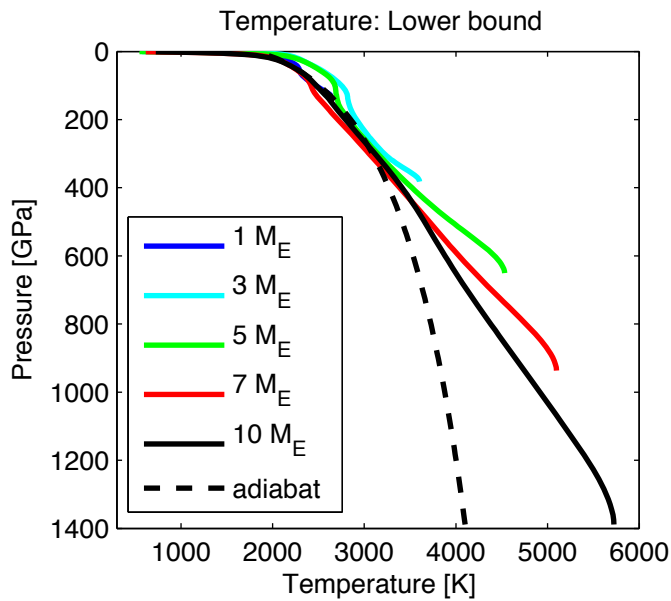
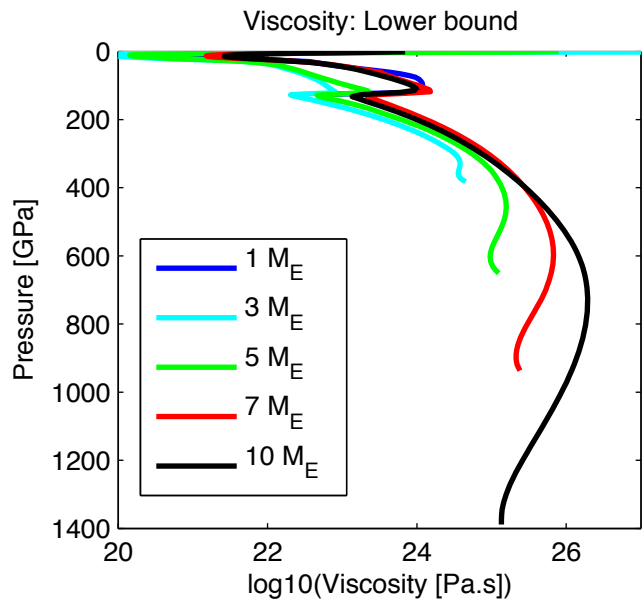